%% file: apstemplate.tex
\documentclass[aps,prl,reprint,groupedaddress]{revtex4-2}
\usepackage[bitstream-charter]{mathdesign}
\usepackage{amsmath}
\usepackage{graphicx}
\usepackage{hyperref}
\hypersetup{colorlinks=true, citecolor=blue, urlcolor=blue, linkcolor=blue}
\usepackage{enumerate}
\usepackage{tabularx} 
\usepackage{booktabs}
\usepackage{xcolor}
\usepackage{xspace}
\usepackage{siunitx}
\usepackage{subfigure}

\begin{document}

\input{sections/prl_title_author_abstract_keyword}

\maketitle

\input{sections/section1}

\input{sections/section2}

\input{sections/section3}

\input{sections/section4}

\input{sections/section5}

\input{sections/acknowledgements}

\bibliographystyle{unsrt}
\bibliography{sections/references.bib}

\end{document}

%% file: sections/prl_title_author_abstract_keyword.tex
\title{Thermal-Gradient Cooling of Atomic Vapor Fluid}


\author{Changhao Cheng}
\author{Jinxian Guo}
\email[]{jxguo@sjtu.edu.cn}

\affiliation{School of Physics and Astronomy, Shanghai Jiao Tong University, Shanghai 200240, China}


\begin{abstract}
The pursuit of high optical depth and long coherence time in atomic ensembles faces a fundamental thermodynamic constraint: heating enhances light-atom coupling via increased density but degrades coherence through thermal broadening, while laser cooling preserves coherence at the cost of density loss. Here, we demonstrate a non-equilibrium strategy that spatially achieves a negative correlation between density and temperature via controlled thermal-gradient transport. By engineering a temperature gradient via laser-cooling in a hot vapor cell, we drive a convective atomic fluid that expels hot atoms at the boundary while confining low-temperature atoms in the central region. 
This dynamic process sustains a density of $n\sim10^{22}\mathrm{m}^{-3}$ and a temperature of tens of kelvins at the center. 
A theoretical scheme based on the Boltzmann-type transport equation is established, which gives Navier-Stokes equations for non-equilibrium thermal-gradient atomic fluid. 
The results of numerical simulation indicate that this scheme can enhance the optical depth while reducing the temperature of the system, establishing a route to bypass equilibrium thermodynamics in room-temperature atom-light interactions, boosting high-performance quantum metrology and quantum information applications.
\end{abstract}


%% file: sections/section1.tex
\section{I. Introduction}
In recent years, due to the simplicity of the experimental setup and high optical depth \cite{pizzey2022laser}, hot atomic systems have achieved successive breakthroughs in quantum information processing \cite{pezze2018quantum} and quantum metrology \cite{giovannetti2011advances}.
In particular, heating an alkali-filled vapor cell provides a straightforward method to increase the atomic number density, thereby enhancing the optical depth (OD). 
For instance, the spin-exchange relaxation-free (SERF) regime, where alkali vapors are heated to extreme temperatures (e.g., 190\textdegree C \cite{Allred2002high}), has dramatically increased atomic density to $n\sim 10^{20}$m$^{-3}$.
This approach effectively improves the interactions between light and atoms, crucial to lowering the standard quantum limit of metrology \cite{wu2023quantum} and enhancing performance of quantum information processing \cite{guo2019high}.

However, such thermally enhanced OD inherently couples with detrimental thermal effects\cite{demtroder1973laser}. 
As the temperature increases, the distribution of atomic velocity broadens due to Doppler effects ($\Delta \nu_D \propto \sqrt{T}$) \cite{Goldenstein2017}, while interatomic collisions induce pressure broadening ($\Delta \nu_C \propto T^{n}$ with $n>0$ varying with atoms) \cite{kluttz2013pressure}.
These mechanisms collectively degrade the coherence time $T_2 \propto 1/\Delta \nu$, imposing a fundamental trade-off between atomic density and coherence. 

At the opposite extreme, laser-cooled atoms trapped in optical lattices or magnet-optical traps achieve sub-Doppler temperatures ($<100 \mathrm{\mu K}$) with coherence times exceeding seconds \cite{bloch2008many}. 
However, even for very high-density cold atomic systems, the atomic number density is limited to $n\sim 10^{18}$m$^{-3}$ \cite{radwell2013cold}. 
This low-density constraint fundamentally limits achievable atom-light coupling and the sensitivity of metrology.

The dichotomy between hot and cold atomic systems underscores a fundamental challenge: achieving high OD with low temperature simultaneously, which significantly restricts the further application of atomic ensembles.
To overcome this challenge, we propose a thermal-gradient cooling strategy that achieves a negative correlation between atomic density and temperature through the control of thermal transport and light-matter interactions. 

The core innovation lies in engineering a controlled non-equilibrium state where a boundary-heating reservoir and velocity-selective laser forces create a radial temperature gradient, actively driving atomic fluid for high atomic density in the cell and lowering the temperature.
When we locally heat the vapor cell periphery, a thermal convection that preferentially expels atoms at the boundary. 
The cooling laser gradually decelerates the atoms and accumulates cold atoms at the center of the vapor cell. 
The two effects create a continuous thermal convection from boundary to the center which greatly increases the atomic density and with a low temperature.
The resulting non-equilibrium steady state achieves a coexistence of atomic densities exceeding $10^{22}\mathrm{m}^{-3}$ (surpassing heated SERF vapors) and low temperatures of tens of  kelvins(far below room temperature). 
This strategy may effectively bridging the performance gap between conventional hot and cold atomic systems, boosting high-performance quantum-enhanced metrology and quantum information applications.

%% file: sections/section2.tex
\section{II. Navier-Stokes Equations for Hot Atoms Decelerated by Laser}
\begin{figure}[htbp]
    \centering
     \subfigure[]{
        \includegraphics{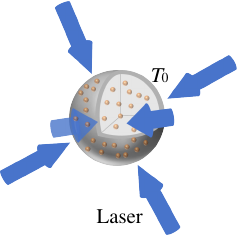}
        \label{fig:1(a)}
    }
    \hfill
     \subfigure[]{
        \includegraphics{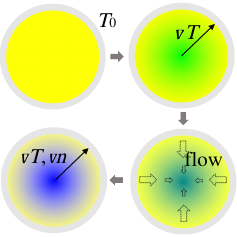}
        \label{fig:1(b)}
    }    
    \caption{Schematic diagram of thermal-gradient cooling. (a) Six Doppler lasers are applied to the vapor cell. The walls of the atomic cell are heated and maintained at a temperature of $T_0$. (b) Evolution of macroscopic quantities on a two-dimensional cross-section: when atoms in a vapor cell are cooled, those near the inner wall maintain boundary temperature, forming an initial temperature gradient. This gradient drives thermal convection currents from the boundary to the center, causing continuous evaporation of liquid atoms from the wall into the interior and establishing a density gradient.}
    \label{fig:1}
\end{figure}

In this work, we consider a spherical atomic cell containing a liquid alkali metal, such as rubidium, which is heated by an external constant thermal source, as shown in Fig. \ref{fig:1(a)}. 
Due to the saturated vapor pressure, most atoms remain in the liquid phase on the inner walls of the cell, while a fraction evaporates into the vapor phase and moves freely within the cell. 
Then, we employ a laser cooling scheme involving six counter-propagating laser beams with detuning $\delta$ and power characterized by the Rabi frequency $\Omega$. 
The cooling lasers induce velocity-dependent momentum dissipation of atomic thermal motion. 
Within the framework of semi-classical theory, the six Doppler lasers generated a cooling force as
\begin{equation}
    \boldsymbol{F}(\boldsymbol{v}) = f(v_x)\boldsymbol{e_x} + f(v_y)\boldsymbol{e_y} + f(v_z)\boldsymbol{e_z},
    \label{3D force}
\end{equation}
where $f$ has the form\cite{balykin2000electromagnetic}
\begin{equation}
\begin{split}
    f(v_i) = \frac{\hbar k \Gamma \Omega^2}{4}[\frac{1}{(\delta + kv_i)^2 + \Gamma^2/4 + \Omega^2/2}\\
    - \frac{1}{(\delta - kv_i)^2 + \Gamma^2/4 + \Omega^2/2}].
\end{split}
\label{1D force}
\end{equation}
Here $k$ is the wave vector of the laser beams, $\Gamma$ is the decay rate of the atom.
Without loss of generality, we assume that the cooling lasers are all plane waves, thus $\nabla\Omega = 0$, and only the scattering force remains.
Under the action of the dissipative force in Eq. \ref{3D force}, the hot atomic ensemble will generate thermal convection through the temperature gradient and then establish a density gradient(Fig. \ref{fig:1(b)}). 
In the following, we establish the Boltzmann-type transport equation to discuss the progress in detail.

\subsection{A. Boltzmann-Type Transport Equation}
High temperature of the cell causes the de Broglie wavelength of the hot alkali atomic vapor to be much smaller than the mean free path $\Bar{\lambda}$, so the two-body collision is dominant among the interactions. 
The Boltzmann transport equation is considered to be able to describe the non-equilibrium dynamics of most monatomic gases dominated by two-body collisions very well. 
The general form of Boltzmann transport equation is \cite{garzo1990influence,perez2014influence,garzo2013kinetic}:
\begin{equation}
    \frac{\partial\omega(\boldsymbol{v}, \boldsymbol{r}, t)}{\partial t} + \boldsymbol{v}\nabla \omega(\boldsymbol{v}, \boldsymbol{r}, t) + \frac{1}{m}\frac{\partial}{\partial \boldsymbol{v}}\boldsymbol{F}(\boldsymbol{v})\omega(\boldsymbol{v},\boldsymbol{r},t) = J[\omega, \omega]
    \label{BTE0}
\end{equation}
when the external force is related to the velocity. 
Here $\omega$ is the velocity distribution function, $\boldsymbol{F}$ indicates the external force and $J[\omega, \omega] = (\frac{\partial \omega}{\partial t})_{collision}$ is a complex integral term representing two-body interaction. Due to the atom-light interaction, the heating effect caused by the spontaneous emission will bring a diffusion of the atomic vapor, describing by the diffusion term $D_{ii}$\cite{balykin2000electromagnetic}: 
\begin{subequations}\label{Dii}
\begin{gather}
    D_{ii} = \frac{\hbar^2k^2\Gamma\Omega^2}{8}\sum_{j = x, y, z}[\frac{\chi_{iij}}{(\delta - \boldsymbol{k}_j\cdot\boldsymbol{v})^2 + \Gamma^2/4 + \Omega^2/2}\nonumber\\ + \frac{\chi_{iij}}{(\delta + \boldsymbol{k}_j\cdot\boldsymbol{v})^2 + \Gamma^2/4 + \Omega^2/2}],\\
    \chi_{iij} = \frac{1}{3} + \delta_{ij}(1 + d_j),\\
    d_j = \frac{[(\delta - \boldsymbol{k}_j\cdot\boldsymbol{v})^2 - 3\Gamma^2/4]\Omega^2}{2[(\delta - \boldsymbol{k}_j\cdot\boldsymbol{v})^2 + \Gamma^2/4 + \Omega^2/2]^2},
\end{gather}
\end{subequations}
here $\boldsymbol{k}_j = k\boldsymbol{e_j}$ with $i,j=x,y,z$. 
With the diffusion modification, the full Boltzmann-type transport equation should be
\begin{equation}
\begin{split}
    \frac{\partial\omega(\boldsymbol{v}, \boldsymbol{r}, t)}{\partial t} + \boldsymbol{v}\nabla \omega(\boldsymbol{v}, \boldsymbol{r}, t) + \frac{1}{m}\frac{\partial}{\partial \boldsymbol{v}}\cdot \boldsymbol{F}(\boldsymbol{v})\omega(\boldsymbol{v}, \boldsymbol{r}, t)\\
    - \frac{1}{m^2}\frac{\partial^2}{\partial\boldsymbol{v}\partial\boldsymbol{v}}: D(\boldsymbol{v})\omega(\boldsymbol{v}, \boldsymbol{r}, t) = J[\omega, \omega]. 
    \label{BTE}
\end{split}
\end{equation}

Compared to the Fokker-Planck type dynamic equation\cite{stenholm1986semiclassical}, which is widely used in decelerating atoms by laser, Eq. \ref{BTE} considers inter-atom collision, which is more suitable to describe the laser-atom interaction in the case of high atomic density.

Considering that the interaction between monatomic gases is mainly elastic collisions and that the influence of inelastic collisional de-coherence on the outer state is insignificant, then for conserved quantities $\psi_{\beta} = m, m\boldsymbol{v}, \frac{m\boldsymbol{u}^2}{2}$(here $\boldsymbol{u} = \boldsymbol{v} - \boldsymbol{c}$ is the velocity relative to the flow velocity), we have:
\begin{equation}
    \int J[\omega, \omega]\psi_{\beta}d\boldsymbol{v} = 0.
    \label{J}
\end{equation}

With the conserved quantities induced equation, we can derive three macroscopic fluid mechanics quantities of atomic density $\rho$, atomic flow velocity $\boldsymbol{c}$ and thermal energy $U$ which are defined as
\begin{equation}
\rho = m\int\omega d\boldsymbol{v}, \boldsymbol{c} = \int \frac{m\boldsymbol{v}}{\rho}\omega d\boldsymbol{v}, U = \int\frac{m\boldsymbol{u}^2}{2\rho}\omega d\boldsymbol{v},
\label{3mac}
\end{equation}
here exists a relationship $U = \frac{3k_BT}{2m}$ between energy density and temperature.
By performing the same integration operation in Eq. \ref{J} on both sides of Eq. \ref{BTE}, the macroscopic fluid mechanics equations can be obtained as
\begin{subequations}\label{mac equations}
\begin{gather}
    \frac{\partial\rho}{\partial t} + \nabla\cdot \rho\boldsymbol{c} = 0,\\
    \frac{\partial}{\partial t}\rho\boldsymbol{c} + \nabla(\rho\boldsymbol{c}\boldsymbol{c} + \overset{\rightleftharpoons}{P}) - \boldsymbol{F_{mac}} = 0,\\
    \frac{\partial}{\partial t}\rho U + \nabla\cdot(\rho U\boldsymbol{c} + \boldsymbol{q}) + \overset{\rightleftharpoons}{P}:\nabla\boldsymbol{c}
        = Q - \boldsymbol{c}\boldsymbol{F_{mac}}.
\end{gather}
\end{subequations}
The stress tensor $P_{ij}$ and heat flux vector $\boldsymbol{q}$ are respectively
\begin{equation}
   P_{ij}(\boldsymbol{r}, t) = \int mu_iu_j\omega d\boldsymbol{v}, \boldsymbol{q} = \int\frac{mu^2\boldsymbol{u}}{2}\omega d\boldsymbol{v},
   \label{P and q}
\end{equation}
where $i,j=x,y,z$. The moments generated by the laser represent the macroscopic force $\boldsymbol{F_{mac}}$ and energy dissipation $Q$ are
\begin{equation}
    \boldsymbol{F_{mac}} = \int \boldsymbol{F}\omega d\boldsymbol{v}, Q = \int(\boldsymbol{F}\boldsymbol{v} + \sum\frac{D_{ii}}{m})\omega d\boldsymbol{v}.
    \label{F and Q}
\end{equation}

Although Eqs. \ref{mac equations} describe the macroscopic fluid dynamics, the macroscopic quantities contained in Eqs. \ref{P and q} and \ref{F and Q} remain unknown until the velocity distribution is obtained. 
Therefore, we return to the Boltzmann-type equation \ref{BTE} to solve for $\omega$, which allows us to concretely derive the Navier-Stokes equations from the macroscopic fluid mechanics equations.

\subsection{B. Navier-Stokes Equations by Chapman-Enskog Method}
Thermal convection typically occurs in the case of a continuous fluid($\Bar{\lambda} \ll L$ where $\lambda$ is the mean free path of atoms and $L$ is the size of the system), so we can apply the series expansion method to solve Eq. \ref{BTE}\cite{cercignani1988boltzmann}. 
According to the Chapman-Enskog method\cite{chapman1990mathematical}, we can expand the velocity distribution function and its derivative with respect to time in a series of powers of $\epsilon$, where $\epsilon$ is the gradient of the flow field. Thus, we have:
\begin{subequations}\label{chapmanenskog}
    \begin{gather}
        \omega = \sum_{i = 0}^{\infty}\epsilon^{i}\omega^{(i)},\\
        \frac{\partial\omega^{(j)}}{\partial t} = \sum_{i = 0}^{\infty}\epsilon^i\frac{\partial_i\omega^{(j)}}{\partial t},
    \end{gather}
\end{subequations}
Chapman-Enskog method also requiress that:
\begin{equation}
    \int\psi_{\beta}\omega^{(n)}d\boldsymbol{v} = 0,n \geq 1.
    \label{cerequire}
\end{equation}

Now Eq. \ref{BTE} can be rewritten as
\begin{equation}
\begin{split}
    \frac{\partial\omega(\boldsymbol{v}, \boldsymbol{r}, t)}{\partial t} + \boldsymbol{v}\nabla \omega(\boldsymbol{v}, \boldsymbol{r}, t) + \frac{1}{m}\frac{\partial}{\partial \boldsymbol{v}}\cdot \boldsymbol{F}(\boldsymbol{v})\omega(\boldsymbol{v}, \boldsymbol{r}, t)\\
    - \frac{1}{m^2}\frac{\partial^2}{\partial\boldsymbol{v}\partial\boldsymbol{v}}: D(\boldsymbol{v})\omega(\boldsymbol{v}, \boldsymbol{r}, t) = \frac{C[\omega, \omega]}{\epsilon},
    \label{BTE2}
\end{split}
\end{equation}
where $C[\omega, \omega] = \epsilon J[\omega, \omega]$ has the same order of magnitude with the left side of Eq. \ref{BTE2}. Substituting Eqs. \ref{chapmanenskog} into Eq. \ref{BTE2}, we obtain the approximate equations at each order:
\begin{subequations}\label{C00}
    \begin{gather}
        C[\omega^{(0)}, \omega^{(0)}] = 0,\\
        \sum_{i = 0}^{j}C[\omega^{(i)}, \omega^{(j - i)}] = \sum_{i = 0}^{j - 1}\frac{\partial_{i}\omega^{(j - i - 1)}}{\partial t} + \boldsymbol{v}\nabla\omega^{(j - 1)}\nonumber\\
    + \frac{1}{m}\frac{\partial}{\partial\boldsymbol{v}}\cdot\boldsymbol{F}\omega^{(j - 1)} - \frac{1}{m^2}\frac{\partial}{\partial\boldsymbol{v}\partial\boldsymbol{v}}:D\omega^{(j - 1)}, j = 1, 2, \cdots.
    \end{gather}
\end{subequations}

The primary solution remains a Maxwellian\cite{cercignani1988boltzmann}:
\begin{equation}
    \omega^{(0)} = n(\frac{m}{2\pi k_BT})^{3/2}\mathrm{exp}[-\frac{m\boldsymbol{u}^2}{2k_BT}].
    \label{omega0}
\end{equation}
By substituting the primary solution $\omega^{(0)}$ into Eqs. \ref{mac equations} we can obtain the zeroth-order Euler equations as
\begin{subequations}\label{eulerequations}
    \begin{gather}
        \frac{\partial_0\rho}{\partial t} = -\nabla\cdot\rho\boldsymbol{c},\\
        \frac{\partial_0 c_i}{\partial t} = -c_j\frac{\partial c_j}{\partial x_j} - \frac{k_B}{\rho}\frac{\partial (nT)}{\partial x_j} + \frac{F_{macj}^{(0)}}{\rho},\\
        \frac{\partial_0 T}{\partial t} = -\boldsymbol{c}\nabla T -\frac{2}{3}(\nabla\cdot\boldsymbol{c})T + \frac{2}{3}\frac{Q^{(0)}}{k_Bn} -\frac{2\boldsymbol{F_{mac}^{(0)}}\boldsymbol{c}}{3k_Bn}.
    \end{gather}
\end{subequations}
For the high-order equations, we let $\omega^{(j)} = \omega^{(0)}h^{(j)}, j = 1, 2, \cdots$ for simplicity. When $j = 1$ we have:
\begin{equation}\label{c1}
\begin{split}
    \omega^{(0)}\mathcal{L}h^{(1)} = \frac{\partial_0\omega^{(0)}}{\partial t} + \boldsymbol{v}\nabla\omega^{(0)}
    + \frac{1}{m}\frac{\partial}{\partial\boldsymbol{v}}\cdot\boldsymbol{F}\omega^{(0)}\\ - \frac{1}{m^2}\frac{\partial}{\partial\boldsymbol{v}\partial\boldsymbol{v}}:D\omega^{(0)}.
\end{split}
\end{equation}
Here we define a linear collision operator $\mathcal{L}$ \cite{cercignani1988boltzmann} for the left side of equation as $\omega^{(0)}\mathcal{L}h^{(1)}=C[\omega^{(0)}, \omega^{(0)}h^{(1)}] + C[\omega^{(0)}h^{(1)}, \omega^{(0)}]$.. Substituting Eqs. \ref{eulerequations} into the right side of Eq. \ref{c1}, we obtain:
\begin{gather}\label{Lh1}
        \mathcal{L}h^{(1)} = \frac{1}{T}(\frac{m}{2k_BT}u^2 - \frac{5}{2})\boldsymbol{u}\nabla T \nonumber\\ + \frac{m}{k_BT}\frac{\partial c_j}{\partial x_i}(u_iu_j - \frac{1}{3}\delta_{ij}u^2) + G(\boldsymbol{u}, T).
\end{gather}
where $G(\boldsymbol{u}, T)$ is the term induced by the external force:
\begin{gather}\label{Gu}
    G(\boldsymbol{u}, T) = \frac{1}{T}(\frac{m}{2k_BT}u^2 - \frac{3}{2})(\frac{2}{3}\frac{Q^{(0)}}{k_Bn} -\frac{2\boldsymbol{F_{mac}^{(0)}}\boldsymbol{c}}{3k_Bn})\nonumber\\ + \boldsymbol{u}\frac{\boldsymbol{F_{mac}^{(0)}}}{k_BTn} + \frac{1}{m}\frac{\partial}{\partial\boldsymbol{v}}\cdot\boldsymbol{F} - \frac{\boldsymbol{F}\boldsymbol{u}}{k_BT}\nonumber\\
    - \frac{1}{m^2}
    \sum[\frac{\partial^2D_{ii}}{\partial v_i^2} - \frac{2m}{k_BT}\frac{\partial D_{ii}}{\partial v_i}u_i + D_{ii}(\frac{m^2u_i^2}{k_B^2T^2} - \frac{m}{k_BT})].
\end{gather}
This is the major difference between our model and conventional BTE , where the external force associated with velocity contributes to the first-order equation. 

Considering the Hermite operator $\mathcal{L}$, we can obtain the solution for $h^{(1)}$ as 
\begin{equation}
\begin{split}
    h^{(1)} = \frac{m}{k_BT}\frac{\partial c_j}{\partial x_i}(u_iu_j - \frac{1}{3}u_ku_k\delta_{ij})\mathcal{A}(\rho, T, u_lu_l) \\
    + \frac{1}{T}\frac{\partial T}{\partial x_i}u_i\mathcal{B}(\rho, T, u_ku_k) + \mathcal{C},
    \label{h1}
\end{split}
\end{equation}
where $\mathcal{A}$, $\mathcal{B}$ still follow the results of the conventional BTE\cite{cercignani1988boltzmann}:
\begin{subequations}\label{solution}
    \begin{gather}
        \mathcal{L}([u_iu_j - \frac{1}{3}u_ku_k\delta_{ij}]\mathcal{A}) = u_iu_j - \frac{1}{3}u_ku_k\delta_{ij},\label{so1}\\
        \mathcal{L}(u_i\mathcal{B}) = u_i(\frac{m}{2k_BT}u_ku_k - \frac{5}{2}),\label{so2}
    \end{gather}
\end{subequations}
while $\mathcal{C}$ satisfies:
\begin{equation}
    \mathcal{LC} = G.
    \label{so4}
\end{equation}
Note that in $h^{(1)}$, apart from the first term itself being orthogonal to $\psi_{\beta}$, the Chapman-Enskog method also imposes the same requirement on $\mathcal{B}$ and $\mathcal{C}$ for satisfying Eq. \ref{cerequire}. At this point, the contributions of the pressure tensor $P_{ij}^{(1)}$, heat flux vector $\boldsymbol{q}^{(1)}$, macroscopic force $\boldsymbol{F_{mac}^{(1)}}$ and energy dissipation $Q^{(1)}$ brought by the first-order solution can be calculated by substituting solution of $h^{(1)}$ in Eqs. \ref{P and q} and \ref{F and Q}.

Then we consider the second-order case with $j = 2$, which gives
\begin{equation}
\begin{split}
    C[\omega^{(0)}, \omega^{(2)}] + C[\omega^{(2)}, \omega^{(0)}] = \frac{\partial_0\omega^{(1)}}{\partial t} + \frac{\partial_1\omega^{(0)}}{\partial t} + \boldsymbol{v}\nabla\omega^{(1)}\\
    + \frac{1}{m}\frac{\partial}{\partial\boldsymbol{v}}\cdot\boldsymbol{F}\omega^{(1)} - \frac{1}{m^2}\frac{\partial}{\partial\boldsymbol{v}\partial\boldsymbol{v}}:D\omega^{(1)} - C[\omega^{(1)}, \omega^{(1)}].   
\end{split}
\label{C02}
\end{equation}
By multiplying $\psi_{\beta}$ with Eq. \ref{C02} and performing the integration along $\boldsymbol{v}$, we can obtain the corresponding fluid mechanics equations as
\begin{subequations}\label{2fluidequations}
    \begin{gather}
        \frac{\partial_1\rho}{\partial t} = 0,\\
        \frac{\partial_1 c_i}{\partial t} = -\frac{1}{\rho}\frac{\partial}{\partial x_j}P_{ij}^{(1)} + \frac{1}{\rho}F_{maci}^{(1)},\\
        \frac{\partial_1 T}{\partial t} = -\frac{2}{3k_Bn}\nabla\boldsymbol{q}^{(1)} - \frac{2}{3k_Bn}P_{ij}^{(1)}\frac{\partial c_j}{\partial x_i} + \frac{2}{3}\frac{Q^{(1)}}{k_Bn} -\frac{2\boldsymbol{F_{mac}^{(1)}}\boldsymbol{c}}{3k_Bn}.
    \end{gather}
\end{subequations}
Adding the Eqs. \ref{2fluidequations} with Eqs. \ref{eulerequations} and setting the factor of series $\epsilon \to 1$, we can obtain:
\begin{subequations}\label{NS equations}
    \begin{gather}
        \frac{\partial\rho}{\partial t} = -\nabla\cdot\rho\boldsymbol{c},\label{ns1} \\
        \frac{\partial c_i}{\partial t} = -c_j\frac{\partial c_j}{\partial x_j} - \frac{k_B}{\rho}\frac{\partial (nT)}{\partial x_i} + \frac{F_{maci}^{(0)} + F_{maci}^{(1)}}{\rho} - \frac{1}{\rho}\frac{\partial}{\partial x_j}P_{ij}^{(1)},\label{ns2} \\
        \frac{\partial T}{\partial t} = -\boldsymbol{c}\nabla T - \frac{2}{3}(\nabla\cdot\boldsymbol{c})T + \frac{2}{3}\frac{Q^{(0)} + Q^{(1)}}{k_Bn} \nonumber\\
        -\frac{2(\boldsymbol{F_{mac}^{(0)}} + F_{mac}^{(1)})\boldsymbol{c}}{3k_Bn}
        - \frac{2}{3k_Bn}\nabla\cdot\boldsymbol{q}^{(1)} - \frac{2}{3k_Bn}P_{ij}^{(1)}\frac{\partial c_j}{\partial x_i},\label{ns3}
    \end{gather}
\end{subequations}
which are the Navier-Stokes equations for a hot atomic ensemble decelerated by laser. 

Before solving the Navier-Stokes equations, we should firstly discuss the boundary conditions of them. 
In our case, we used a spherical vapor cell filled with atomic vapor fluid. Due to the stability of the vapor pressure near the inner wall and the high symmetry of the system, it is easy for us to obtain:
\begin{subequations}\label{boundary conditions}
    \begin{gather}
        \rho|_{\boldsymbol{r} \in \Sigma} = \rho_0, T|_{\boldsymbol{r} \in \Sigma} = T_0, c|_{\boldsymbol{r} = 0} = 0, \label{bc0}\\
        \nabla\rho|_{\boldsymbol{r} = 0} = 0, \nabla\cdot\boldsymbol{c}|_{\boldsymbol{r} = 0} = 0,\nabla T|_{\boldsymbol{r} = 0} = 0. \label{bc1}
    \end{gather}
\end{subequations}

%% file: sections/section3.tex
\section{III. Stationary Heat Conduction Equation}
Solving the non-steady-state Navier-Stokes equations is usually complex. Therefore, we limit ourselves to solving the stationary situation of Eqs. \ref{NS equations} when the size of the vapor cell is not too large and the dissipation caused by laser cooling is not too strong.

At steady state, Eq. \ref{ns1} and Eq. \ref{bc0} give
\begin{equation}
    \boldsymbol{c} \equiv 0.
    \label{cequ0}
\end{equation}
It means that a stationary state of the atomic system must be no particle flow inside the cell, which is easy to imagine in a closed system. In this case, $\boldsymbol{F_{mac}^{(0)}} = 0$, and $G$ can be simplified to
\begin{equation}\label{Gsimple}
\begin{split}
    G(\boldsymbol{v}) = \frac{1}{T}(\frac{m}{2k_BT}v^2 - \frac{3}{2})\frac{2}{3}\frac{Q^{(0)}}{k_Bn} + \frac{1}{m}\frac{\partial}{\partial\boldsymbol{v}}\cdot\boldsymbol{F} - \frac{\boldsymbol{F}\boldsymbol{v}}{k_BT}\\
     - \frac{1}{m^2}\sum[\frac{\partial^2D_{ii}}{\partial v_i^2} - \frac{2m}{k_BT}\frac{\partial D_{ii}}{\partial v_i}v_i + D_{ii}(\frac{m^2v_i^2}{k_B^2T^2} - \frac{m}{k_BT})].
\end{split}
\end{equation}
Now $G$ becomes an even function.

Returning to the expression of $h^{(1)}$(Eq.  \ref{h1}), since there is no flow, the first term naturally vanishes. We can continue to simplify the first-order macroscopic quantities covered by Eqs. \ref{P and q} and \ref{F and Q} under such circumstances. As $G$ is an even function, it is easy to obtain that $\mathcal{C}$ is also an even function. Therefore, the inner product of any odd function (such as $v^2\boldsymbol{v}$ and $\boldsymbol{F}$) with $\mathcal{C}$ will become 0. Finally, considering the symmetry of laser cooling in the three directions and  $\int v^2\mathcal{C}\omega^{(0)}d\boldsymbol{v} = 0$ indicated by Eq. \ref{cerequire}, there is also $\int v_i^2\mathcal{C}\omega^{(0)}d\boldsymbol{v} = 0$. Thus, we obtain the simplified first-order form of the macroscopic quantities:
\begin{subequations}\label{1thso}
\begin{gather}
    P_{ij}^{(1)} = 0, \label{P1}\\
    F_{mac,i}^{(1)} = \frac{1}{T}\frac{\partial T}{\partial x_i}\int F_iv_i\mathcal{B}\omega^{(0)}d\boldsymbol{v}, \label{F1}\\
    {q}_{i}^{(1)} = \frac{m}{6T}\frac{\partial T}{\partial x_i}\int v^4\mathcal{B}\omega^{(0)}d\boldsymbol{v}, \label{q1}\\
    Q^{(1)} = \int\boldsymbol{F}\boldsymbol{v}\mathcal{C}\omega^{(0)}d\boldsymbol{v}. \label{Q1}
\end{gather}
\end{subequations}
According to Fourier's law $\boldsymbol{q} = -\kappa\nabla T$ where $\kappa$ is the thermal conductivity, we can obtain that $\kappa(T) = -\frac{m}{6T}\int v^4\mathcal{B}\omega^{(0)}d\boldsymbol{v}$. Noticing that the six Doppler lasers have the same parameters, we can set
\begin{equation}
    I_i(T) = I(T)= \frac{1}{T}\int F_iv_i\mathcal{B}\omega^{(0)}d\boldsymbol{v}.
    \label{Ii}
\end{equation}
As a result, we can simplify the expression of macroscopic force as
\begin{equation}
    \boldsymbol{F_{mac}^{(1)}} = I(T)\nabla T.
    \label{thermal doppler laser pressure}
\end{equation}
Now the stationary Navier-Stokes equations can be simplified to:
\begin{subequations}\label{simple ns}
    \begin{gather}
        k_B\nabla(nT) = \boldsymbol{F_{mac}^{(1)}}(T), \label{stationary mom equ}\\
        \nabla\cdot\kappa(T)\nabla T + Q^{(0)}(n, T) + Q^{(1)}(T) = 0, \label{stationary ene equ}
    \end{gather}
\end{subequations}
Through Eq. \ref{stationary mom equ}, we can solve for the number density $n$ as a function of temperature $T$:
\begin{equation}
    n(T) = \frac{\int_{T_0}^{T}I(T)dT + n_0k_BT_0}{k_BT}.
\label{nT}
\end{equation}
Substituting the Eq. \ref{nT} into the Eq. \ref{stationary ene equ}, we can obtain the stationary heat conduction equation as
\begin{equation}
    \nabla\cdot\kappa(T)\nabla T + Q^{(0)}(n(T), T) + Q^{(1)}(T) = 0.
    \label{hce}
\end{equation}
By numerically simulating the stationary heat conduction equation, the density distribution and the temperature distribution in the atomic vapor can be obtained.

%% file: sections/section4.tex
\section{IV. Numerical Results}
To solve the Eq. \ref{hce}, an exact form of $\kappa$, $I(T)$, $Q^{(0)}$ and $Q^{(1)}$ should be obtained which requires a specific expression of inter-atom interaction. Generally, rigid spheres is used to model the atoms which gives the potential energy as
\begin{equation}
    V = 
    \begin{cases}
        \infty, & r < \sigma \\
        0, & r \geq \sigma
    \end{cases},
\end{equation}
where $\sigma$ is the average collision radius. 

\begin{table*}[htbp]
    \centering
    \caption{Th Maximum order of magnitude of $|Q^{(1)}/Q^{(0)}|$ estimated by the BGK model within the temperature range from 0 to $T_0$}
    \label{tab:1}
    \begin{tabularx}{\textwidth}{cc *{6}{X}} 
        \toprule
        $T_0$ (K) & $P_0$ (Pa)\cite{steck2001rubidium} & \multicolumn{6}{c}{$|Q^{(1)}/Q^{(0)}|_{\text{MAX}}$ at different $\delta$} \\ 
        \cmidrule(lr){3-8}
        & & $-10\Gamma$ & $-30\Gamma$ & $-50\Gamma$ & $-70\Gamma$ & $-100\Gamma$ & $-150\Gamma$ \\
        \midrule
        450 & 2.19 & $2.0\times10^{-3}$ & $4.1\times10^{-3}$ & $3.7\times10^{-4}$ & $1.1\times10^{-3}$ & $3.6\times10^{-3}$ & $7.5\times10^{-5}$ \\
        500 & 17.29 & $2.6\times10^{-4}$ & $5.2\times10^{-4}$ & $4.7\times10^{-5}$ & $1.4\times10^{-4}$ & $4.5\times10^{-4}$ & $9.5\times10^{-6}$ \\
        550 & 93.84 & $6.9\times10^{-5}$ & $9.6\times10^{-5}$ & $8.7\times10^{-6}$ & $2.6\times10^{-5}$ & $8.4\times10^{-5}$ & $1.8\times10^{-6}$ \\
        \bottomrule
    \end{tabularx}
\end{table*}

Under the rigid spheres approximation, we can expand $\mathcal{B}$ in accordance with the right side of Eq. \ref{so2}, and truncate only to the first order to obtain the approximate result as
\begin{equation}
    \mathcal{B} \approx \mathcal{B}_1L_{1}^{3/2}(\frac{mv^2}{2k_BT}) = \frac{15}{32n\sigma^2}(\frac{m}{\pi k_BT})^{\frac{1}{2}}(\frac{m}{2k_BT}v^2 - \frac{5}{2}),
\end{equation}
where $L_{n}^{\alpha}$ represents the generalized Laguerre polynomial.
Then the thermal conductivity $\kappa(T)$ and $I(T)$ of rigid sphere atoms are
\begin{subequations}\label{kappa and I}
\begin{gather}
    \kappa(T) \approx \frac{75}{64\sigma^2}(\frac{k_B^3T}{\pi m})^{\frac{1}{2}}, \label{kappa}\\
    I(T) \approx \frac{15}{32\sigma^2}(\frac{m}{\pi k_BT^3})^{\frac{1}{2}}\int f(v_i)v_i(\frac{m}{2k_BT}v^2 - \frac{5}{2})p^{(0)}d\boldsymbol{v}, \label{I(rigid sphere)}
\end{gather}
\end{subequations}
here $p^{(0)} = \omega^{(0)}/n$. 

Even when approximated by the rigid-sphere model, $\mathcal{C}$ still takes the form of an infinite series owing to the complexity of $G(\boldsymbol{v})$, which makes the calculation of $Q^{(1)}$ difficult. 
Nonetheless, we are aware that the first-order contribution of heat dissipation is solely related to temperature. Meanwhile, the zeroth-order contribution is proportional to the number density. 
If the number density is large enough, the contribution of $Q^{(1)}$ can be disregarded, which significantly reduces the difficulty of the solution. 

To further prove the contribution of $Q^{(1)}$, we calculate the approximate value of $Q^{(1)}/Q^{(0)}$ with the Bhatnagar-Gross-Krook (BGK) model \cite{bhatnagar1954model}, which has found extensive applications in solving the BTE. Table \ref{tab:1} presents the order of magnitude of $Q^{(1)}$ with $\Omega = 10\Gamma$. It can be seen that $Q^{(1)}$ contributes little and can be disregarded when the temperature is high enough. In this manner, the heat conduction equation \ref{hce} is simplified, enabling us to easily derive the stationary temperature and number density distribution. 

\begin{figure*}[htbp]
    \centering
     \subfigure[]{
        \includegraphics{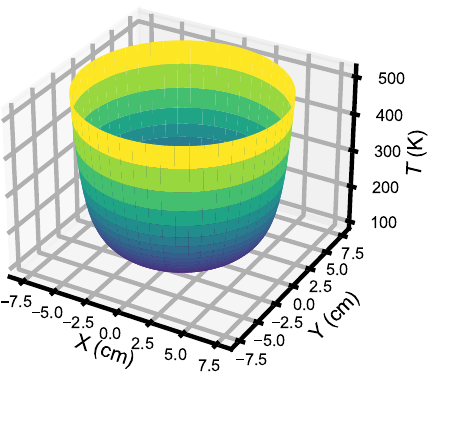}
        \label{fig:2(a)}
    }
    \hfill
     \subfigure[]{
        \includegraphics{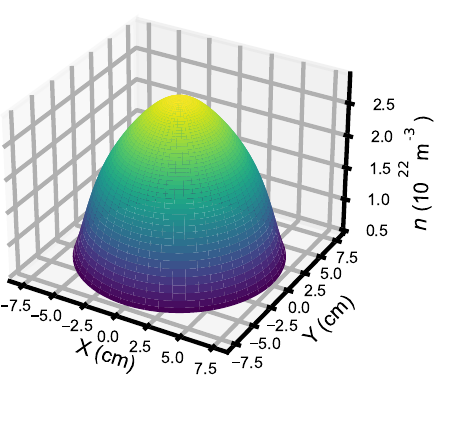}
        \label{fig:2(b)}
    }    
    \caption{Stationary temperature distribution and number density distribution of thermal gradient cooling on the cross-section of the vapor cell ($R = 7.5cm, T_0 = 520\mathrm{K}, \Omega = 10\Gamma, \delta = -100\Gamma$). }
    \label{fig:2}
\end{figure*}

By numerically simulating the Eq. \ref{hce} under rigid spheres approximation, we plot the temperature and number density distribution on the cross-section passing through the center of the sphere cell with a radius of 7.5cm on Fig. \ref{fig:2}.  Here we set the temperature of the thermal reservoir as $T_0 = 520\mathrm{K}$ and specify the parameters of the cooling laser as $\Omega = 10\Gamma$ and $\delta = -100\Gamma$. It can be seen that the temperature decreases rapidly from $520\mathrm{K}$ at the boundary to 88.5K at the center ($X=0 cm$, $Y=0 cm$) as the distance to the center decreases, with a minimum value occurring at the center. 
This indicates that we have indeed established a thermal gradient in the vapor cell through the heat reservoir and laser cooling. This temperature gradient ensures the low temperature at the center of the vapor cell. 
Meanwhile, as can be seen from Fig. \ref{fig:2(b)}, this temperature gradient also creates a gradient in the atomic density distribution, with the density increasing rapidly from $4.9\times10^{21}\mathrm{m}^{-3}$ at the boundary to $2.9\times10^{22}\mathrm{m}^{-3}$ at the center, thereby achieving an increase in atomic density. 
The results prove that our approach can indeed achieve both high average density and low average temperature of the entire atomic cell, as well as the temperature and density at the center. 

In this case, if a probe laser passes through the center, the effective optical depth increases to 4.3 folds of the heating-only method. To achieve such an optical depth with a conventional heating scheme, the temperature must be raised to 569K. We can consider that under thermal-gradient cooling, the atomic vapor fluid achieves a negative correlation between optical depth and temperature.

\begin{figure*}[htbp]
    \centering
     \subfigure[]{
        \includegraphics{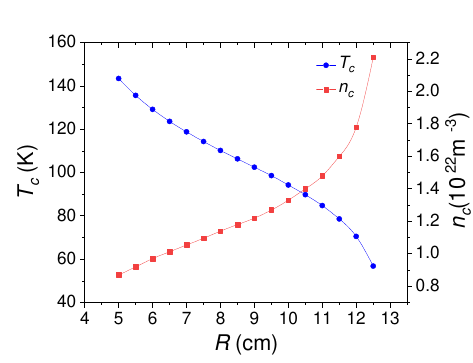}
        \label{fig:3(a)}
    }
    \hfill
     \subfigure[]{
        \includegraphics{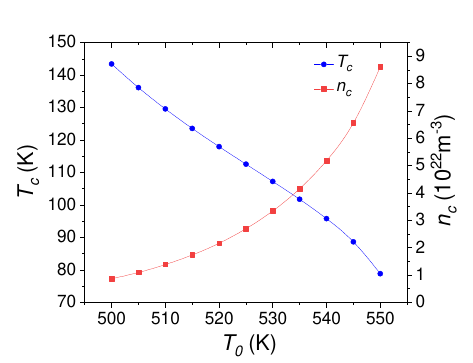}
        \label{fig:3(b)}
    }    
    \caption{The central temperature $T_c$ and number density $n_c$($\Omega = 10\Gamma,\delta = -100\Gamma$) as a function of (a) the radius of the vapor cell $R$($T_0 = 500$K) and (b) the boundary temperature $T_0$($R = 5$cm).}
    \label{fig:3}
\end{figure*}

To further enhance the performance of our strategy, we investigate the central temperature $T_c$ varying with the cell size $R$ and the temperature of thermal reservoir $T_0$, as depicted in Fig. \ref{fig:3}. 
It presents that the central temperature decreases with the growing size of the vapor cell. The larger the size, the lower the temperature.  As a comparison, the atomic density increases with the growing size of the vapor cell.
It is quite evident that the expansion of the vapor cell lead to the central atoms less affected by the heat reservoir, thereby achieving lower temperatures and higher densities.
When the boundary temperature $T_0$ is increased, there is a substantial rise in the number density of gaseous atoms. 
An interesting observation is that the central temperature $T_c$ decrease with the increase in boundary temperature $T_0$.
This is because that the dissipation term is proportional to the number density, consequently increasing with the rising number density. 
The enhanced dissipation effect caused by increasing number density even beat the heating effect of the increasing boundary temperature $T_0$.
Hence, within a specific temperature range, increasing the boundary temperature can enhance the cooling effect. 

%% file: sections/section5.tex
\section{V. Conclusion and Discussion}

In this paper, by establishing the Boltzmann-type transport equation of the interaction between light and continuous atomic vapor flow, the Navier-Stokes equations were obtained. The steady-state results indicate that the thermal-gradient cooling scheme can increase the optical depth while reducing the temperature.

With our strategy, the average collision time of atoms $\tau$ in the cell can be extended. Taking Fig. \ref{fig:2} as an example, we can estimate the time available for experiments. At the center, the average collision time after cooling is $\tau \sim 163 ns$ which allows several quantum coherent operations \cite{guo2019high}. 

It is necessary to emphasize that not all cases have a steady state. When the system size is large enough or the number density of gaseous atoms is high enough, the heat conduction at the boundary may not be sufficient to counteract the huge dissipation generated internally, and Eq. \ref{hce} will not have a stationary solution, one must solve the non-steady-state Navier-Stokes equations \ref{NS equations} to analyse such situation.

Furthermore, this paper considers the situation where atomic vapor can be regarded as continuous fluid without temperature jump at the boundary. This requires a small Knudsen number $K_n = \frac{\Bar{\lambda}}{2R}$ with mean free path $\Bar{\lambda}$ of alkali vapor, a key parameter in the field of fluid mechanics. 
For a spherical rubidium vapor cell with a radius of 5 cm, $K_n\approx 0.1$ when the temperature reaches \qty{160}{\celsius}; while when the temperature reaches \qty{220}{\celsius}, $K_n\approx 0.01$.
This experimental condition can be reached for SERF magnetometers\cite{li2018serf} and experiments on superheated vapor spectroscopy\cite{vadla2006comparison,vadla2006accurate,wells2014alkali}.

%% file: sections/acknowledgements.tex
\section*{Acknowledgments}

This work was supported by the National Natural Science Foundation of China (NSFC) [12374331,123B100015]. During the preparatory phase of this work, we engaged in some fruitful discussions with Hepeng Zhang and ZhenWei Yao from the Institute of Natural Sciences, and Xiongfeng Yang from the School of Mathematical Sciences, all at Shanghai Jiao Tong University. Their insights and perspectives have been highly illuminating and have significantly contributed to the progress of this work. We are sincerely grateful for their valuable contributions. 

%% file: apstemplate.bbl
\begin{thebibliography}{10}

\bibitem{pizzey2022laser}
D~Pizzey, JD~Briscoe, FD~Logue, FS~Ponciano-Ojeda, SA~Wrathmall, and IG~Hughes.
\newblock Laser spectroscopy of hot atomic vapours: from’scope to theoretical fit.
\newblock {\em New Journal of Physics}, 24(12):125001, 2022.

\bibitem{pezze2018quantum}
Luca Pezz\`e, Augusto Smerzi, Markus~K. Oberthaler, Roman Schmied, and Philipp Treutlein.
\newblock Quantum metrology with nonclassical states of atomic ensembles.
\newblock {\em Rev. Mod. Phys.}, 90:035005, Sep 2018.

\bibitem{giovannetti2011advances}
Vittorio Giovannetti, Seth Lloyd, and Lorenzo Maccone.
\newblock Advances in quantum metrology.
\newblock {\em Nature photonics}, 5(4):222--229, 2011.

\bibitem{Allred2002high}
J.~C. Allred, R.~N. Lyman, T.~W. Kornack, and M.~V. Romalis.
\newblock High-sensitivity atomic magnetometer unaffected by spin-exchange relaxation.
\newblock {\em Phys. Rev. Lett.}, 89:130801, Sep 2002.

\bibitem{wu2023quantum}
Shuhe Wu, Guzhi Bao, Jinxian Guo, Jun Chen, Wei Du, Minwei Shi, Peiyu Yang, Liqing Chen, and Weiping Zhang.
\newblock Quantum magnetic gradiometer with entangled twin light beams.
\newblock {\em Science Advances}, 9(15):eadg1760, 2023.

\bibitem{guo2019high}
Jinxian Guo, Xiaotian Feng, Peiyu Yang, Zhifei Yu, L.~Q. Chen, Chun-Hua Yuan, and Weiping Zhang.
\newblock High-performance {Raman} quantum memory with optimal control in room temperature atoms.
\newblock {\em Nature Communications}, 10(1):148, January 2019.

\bibitem{demtroder1973laser}
Wolfgang Demtr{\"o}der.
\newblock {\em Laser spectroscopy}, volume~2.
\newblock Springer, 1973.

\bibitem{Goldenstein2017}
Christopher~S. Goldenstein, R.Mitchell Spearrin, Jay.~B. Jeffries, and Ronald~K. Hanson.
\newblock Infrared laser-absorption sensing for combustion gases.
\newblock {\em Progress in Energy and Combustion Science}, 60:132--176, 2017.

\bibitem{kluttz2013pressure}
Kelly~A. Kluttz, Todd~D. Averett, and Brian~A. Wolin.
\newblock Pressure broadening and frequency shift of the ${D}_{1}$ and ${D}_{2}$ lines of rb and k in the presence of ${}^{3}$he and n${}_{2}$.
\newblock {\em Phys. Rev. A}, 87:032516, Mar 2013.

\bibitem{bloch2008many}
Immanuel Bloch, Jean Dalibard, and Wilhelm Zwerger.
\newblock Many-body physics with ultracold gases.
\newblock {\em Rev. Mod. Phys.}, 80:885--964, Jul 2008.

\bibitem{radwell2013cold}
N~Radwell, G~Walker, and S~Franke-Arnold.
\newblock Cold-atom densities of more than 10 12 cm- 3 in a holographically shaped dark spontaneous-force optical trap.
\newblock {\em Physical Review A—Atomic, Molecular, and Optical Physics}, 88(4):043409, 2013.

\bibitem{balykin2000electromagnetic}
VI~Balykin, VG~Minogin, and VS~Letokhov.
\newblock Electromagnetic trapping of cold atoms.
\newblock {\em Reports on Progress in Physics}, 63(9):1429, 2000.

\bibitem{garzo1990influence}
V~Garz{\'o}, A~Santos, and JJ~Brey.
\newblock Influence of nonconservative external forces on self-diffusion in dilute gases.
\newblock {\em Physica A: Statistical Mechanics and its Applications}, 163(2):651--671, 1990.

\bibitem{perez2014influence}
Jos{\'e}~Carlos P{\'e}rez-Fuentes and Vicente Garz{\'o}.
\newblock Influence of a drag force on linear transport in low-density gases. stability analysis.
\newblock {\em Physica A: Statistical Mechanics and its Applications}, 410:428--438, 2014.

\bibitem{garzo2013kinetic}
Vicente Garz{\'o} and Andr{\'e}s Santos.
\newblock {\em Kinetic theory of gases in shear flows: nonlinear transport}, volume 131.
\newblock Springer Science \& Business Media, 2013.

\bibitem{stenholm1986semiclassical}
Stig Stenholm.
\newblock The semiclassical theory of laser cooling.
\newblock {\em Reviews of modern physics}, 58(3):699, 1986.

\bibitem{cercignani1988boltzmann}
Carlo Cercignani and Carlo Cercignani.
\newblock {\em The boltzmann equation}.
\newblock Springer, 1988.

\bibitem{chapman1990mathematical}
Sydney Chapman and Thomas~George Cowling.
\newblock {\em The mathematical theory of non-uniform gases: an account of the kinetic theory of viscosity, thermal conduction and diffusion in gases}.
\newblock Cambridge university press, 1990.

\bibitem{steck2001rubidium}
Daniel~A Steck.
\newblock Rubidium 87 d line data.
\newblock 2001.

\bibitem{bhatnagar1954model}
Prabhu~Lal Bhatnagar, Eugene~P Gross, and Max Krook.
\newblock A model for collision processes in gases. i. small amplitude processes in charged and neutral one-component systems.
\newblock {\em Physical review}, 94(3):511, 1954.

\bibitem{li2018serf}
Jundi Li, Wei Quan, Binquan Zhou, Zhuo Wang, Jixi Lu, Zhaohui Hu, Gang Liu, and Jiancheng Fang.
\newblock Serf atomic magnetometer--recent advances and applications: A review.
\newblock {\em IEEE Sensors Journal}, 18(20):8198--8207, 2018.

\bibitem{vadla2006comparison}
C~Vadla, Robert Beuc, Vlasta Horvatic, Mladen Movre, Alfred Quentmeier, and Kay Niemax.
\newblock Comparison of theoretical and experimental red and near infrared absorption spectra in overheated potassium vapour.
\newblock {\em The European Physical Journal D-Atomic, Molecular, Optical and Plasma Physics}, 37:37--49, 2006.

\bibitem{vadla2006accurate}
Cedomil Vadla, Vlasta Horvatic, and Kay Niemax.
\newblock Accurate determination of the atomic number density in dense cs vapors by absorption measurements of cs 2 triplet bands.
\newblock {\em Applied Physics B}, 84:523--527, 2006.

\bibitem{wells2014alkali}
N~Wells, T~Driskell, and J~Camparo.
\newblock Alkali pressure shifts and their temperature dependence: Measurements with the rb isoclinic point.
\newblock In {\em 2014 IEEE International Frequency Control Symposium (FCS)}, pages 1--1. IEEE, 2014.

\end{thebibliography}
